\documentclass[12pt]{article}
\usepackage{amssymb,amsmath}

\newtheorem{theorem}{Theorem}[section]

\newtheorem{proposition}[theorem]{Proposition}

\newenvironment{proof}[1][Proof]{\begin{trivlist}
\item[\hskip \labelsep {\bfseries
#1}]}{\end{trivlist}}
\newenvironment{definition}[1][Definition]{\begin{trivlist}
\item[\hskip \labelsep {\bfseries
#1}]}{\end{trivlist}}
\newenvironment{example}[1][Example]{\begin{trivlist} \item[\hskip
\labelsep {\bfseries
#1}]}{\end{trivlist}}
\newenvironment{remark}[1][Remark]{\begin{trivlist}
\item[\hskip
\labelsep {\bfseries #1}]}{\end{trivlist}} \newcommand{\qed}{\nobreak
\ifvmode \relax \else
\ifdim\lastskip<1.5em \hskip-\lastskip
\hskip1.5em plus0em minus0.5em \fi \nobreak \vrule height0.75em width0.5em
depth0.25em\fi}
\newcommand{\E}{\textup{\bf E}}
\def\lh{\hbox to 15pt{\vbox{\vskip 6pt\hrule width 6.5pt height 1pt} \kern
-4.0pt\vrule height 8pt width 1pt\hfil}} 

\begin{document}

\begin{centering}

{\bf \Large Symmetry Reduction of sh-Lie Structures}

{\bf \Large \vspace{.5cm} and of Local Functionals}
\\

\renewcommand{\thefootnote}{\fnsymbol{footnote}}
\vspace{1.4cm}
{\large Samer Al-Ashhab}\\

\vspace{.5cm}

Department of Mathematics, North Carolina State University, Raleigh,
NC 27695-8205.\\
\it E-mail: ssalash@unity.ncsu.edu\\
\rm
\vspace{.5cm}

\rm
\vspace{.5cm}
\vspace{.5cm}

\begin{abstract}
Reduced sh-Lie structures have been studied for the case when
a Lie group acts on the fibers of a vector bundle while preserving the base
space of the bundle. In this paper we investigate how one obtains a reduced
sh-Lie structure using the ideas of symmetry reduction where the action of the
Lie group is transversal to the fibers of the bundle. We also show how local
functionals are reduced using these ideas.
\end{abstract}

\end{centering}

\noindent
{\small Keywords: Reduction, sh-Lie algebra, jet bundle, Poisson bracket. \\
AMS Subject Classification; Primary: 53Z05. Secondary: 16W22.}

\section{Introduction}
One assumes the existence of a ``Poisson
bracket" on the space of local functionals that have the form
$${\cal
P}(\phi)=\int_M(P\circ j^{\infty}\phi)(x) Vol_M$$ where $P:J^{\infty}E
\rightarrow {\bf R}$ is a local function (which means that it is a function on
some finite bundle, i.e., $P:J^pE\rightarrow {\bf R}$ for some finite $p$),
$\pi:E\rightarrow M$ is a vector bundle, $\phi$ is a section of the bundle $\pi$
and
$j^{\infty}\phi$ is the induced section of $J^\infty E$. In other words these
functionals are evaluated at sections $\phi$ of $\pi$.

This bracket is assumed to satisfy the Jacobi identity and so defines a Lie
algebra structure on the space of local functionals ${\cal F}$. On the other
hand there is no obvious commutative multiplication of such functionals and
consequently ${\cal F}$ is not a Poisson algebra. This is such a well-known
development that we may refer to standard monographs on the subject. In
particular we call attention to \cite{D91} and \cite{O86} for classical
expositions and to \cite{HT92} for a quantum field theoretic development.

It was shown in \cite{BFLS98} that a Poisson bracket on the space of local
functionals induces what is known as strongly homotopy Lie structure (sh-Lie
structure) on a part of the variational bicomplex which we refer to as the ``de
Rham complex" on $J^{\infty}E.$ This sh-Lie structure is given by three
mappings $l_1,l_2,$ and $l_3$ defined on this complex. The mapping $l_2$ is
skew-symmetric and bilinear, and it may be regarded as defining a ``bracket"
but one which generally fails to satisfy the Jacobi identity. In fact $l_2$
satisfies the Jacobi identity if $l_3=0.$ In a sense this sh-Lie structure is
an anti-derivative of the Poisson bracket. We refer the interested reader to
\cite{BFLS98}, \cite{LM95}, and \cite{LS93} for more on sh-Lie structures (also
sh-Lie algebras and $L_\infty$ algebras).

In \cite{AF03} ideas related to canonical transformations of these structures
were studied as well as ideas of reduction. In this paper we investigate how
the ideas of symmetry reduction as in \cite{AF97} can be applied to obtain a
reduction of sh-Lie structures and of functionals. This reduction depends on
the existence of a cochain map between a given complex and a ``reduced one". In
fact we can think of this work as an extension of the work done in \cite{AF03},
where in \cite{AF03} reduction was studied when the induced action of a Lie
group $G$ on the base manifold $M$ is the identity map. In this paper, however,
we assume that the action of the Lie group $G$ is transversal to the fibers
where we consider a different kind of reduction. In a sense this work is
complementary to that presented in \cite{AF03}.

After presenting some background material in section 2, we summarize the
conditions for the {\em induced} transformations on the space of local
functionals to be {\em canonical}. In section 3 we recall how these induced
canonical transformations relate to the sh-Lie structure maps. We omit the
proofs of the theorems in sections 2 and 3 and refer the reader to \cite{AF03}.
In section 4 we assume that one has a Lie group acting by canonical
transformations on the space of local functionals. We then show the existence
of an sh-Lie structure on a corresponding reduced complex. There is a brief
discussion of reduction of functionals in section 5.

\section{Bundle automorphisms preserving the Poisson structure on the space of
functionals} \label{BAS}
\subsection{Background material}
In this section we introduce some
of the terminology and concepts that are used in this paper, in addition to
some of the simpler results that will be needed. Our exposition and notation
closely follows that in \cite{BFLS98}. First let $\pi: E \to M$ be a vector
bundle where the base space $M$ is an $n$-dimensional manifold and let
$J^\infty E$ be the infinite jet bundle of $E.$ The restriction of the infinite
jet bundle over  an appropriate open set $U\subset M$ is trivial with fiber an
infinite dimensional vector space $V^\infty$.  The bundle \begin{eqnarray*}
\pi^\infty : J^\infty E_U=U\times V^\infty \rightarrow U \end{eqnarray*} then
has induced coordinates given by \begin{eqnarray*}
(x^i,u^a,u^a_i,u^a_{i_1i_2},\dots,). \end{eqnarray*} We use multi-index notation
and the summation convention throughout the paper. If $j^{\infty}\phi$ is the
section of $J^{\infty}E$ induced by a section $\phi$ of the bundle $\pi$, then
$u^a\circ j^{\infty}\phi=u^a\circ \phi$ and $$u^a_I\circ j^{\infty}\phi=
(\partial_{i_1}\partial_{i_2}...\partial_{i_r})(u^a\circ j^{\infty}\phi)$$
where $r$ is the order of the symmetric multi-index
$I=\{i_1,i_2,...,i_r\}$,with the convention that, for $r=0$, there are no
derivatives. For more details see \cite{A96}, \cite{KV98}, \cite{O86}, and
\cite{S89}.

Let $Loc_E$ denote the algebra of local functions where a local function on
$J^\infty E$ is defined to be the pull-back of a smooth real-valued function on
some finite jet bundle $J^p E$ via the projection from $J^\infty E$ to $J^p E$.
Let $Loc_E^0$ denote the subalgebra of $Loc_E$ such that $P \in Loc_E^0$ iff
$(j^\infty \phi)^* P$ has compact support for all $\phi \in \Gamma E$ with
compact support, and where $\Gamma E$ denotes the set of sections of the bundle
$\pi: E \to M$. The complex of differential forms $\Omega^*(J^{\infty}E,d)$ on
$J^{\infty}E$ possesses a differential ideal, the ideal ${C}$ of contact forms
$\theta$ which satisfy $(j^{\infty}\phi)^* \theta=0$ for all sections $\phi$
with compact support. This ideal is generated by the contact one-forms, which
in local coordinates assume the form $\theta^a_J=du^a_J-u^a_{iJ}dx^i$. Contact
one-forms of order $0$ satisfy $(j^{1}\phi)^*(\theta)=0$, where in local
coordinates they assume the form $\theta^a= du^a-u^a_idx^i$.

Using the contact forms, we see that the complex $\Omega^*(J^{\infty}E,d)$
splits as a bicomplex $\Omega ^{r,s}(J^ \infty E)$ (though the finite level
complexes $\Omega^*(J^pE)$ do not), where $\Omega ^{r,s}(J^ \infty E)$ denotes
the space of differential forms on $J^\infty E$ with $r$ horizontal components
and $s$ vertical components. The bigrading is described by writing a
differential $p$-form $\alpha=\alpha_{IA}^{\bf J}(\theta^A_{\bf J} \wedge
dx^I)$ as an element of $\Omega^{r,s}(J^{\infty}E)$, with $p=r+s$, and
\begin{eqnarray*} dx^I=dx^{i_1}\wedge...\wedge dx^{i_r} \quad {\rm and}
  \quad \theta^A_{\bf J}=\theta^{a_1}_{J_1}\wedge...\wedge
  \theta^{a_s}_{J_s}.
\end{eqnarray*}

Let $C_0$ denote the {\it set of contact one-forms of order zero}, and notice
that both $C_0$ and $\Omega ^{n,1} = \Omega ^{n,1}(J^ \infty E)$ are modules
over $Loc_E$. Let $\Omega ^{n,1}_0$ denote the subspace of $\Omega ^{n,1}$
which is locally generated by the forms $\{(\theta ^a \wedge d^nx)\}$ over
$Loc_E$. We assume the existence of a mapping $\omega$ from $\Omega ^{n,1}_0
\times \Omega ^{n,1}_0$ to $Loc_E$, such that $\omega$ is a skew-symmetric
module homomorphism in each variable separately. In local coordinates let
$\omega ^{ab} = \omega (\theta ^a \wedge \nu,\theta ^b \wedge \nu)$, where
$\nu$ is a volume element on $M$ (notice that in local coordinates $\nu$ takes
the form $\nu = f d^nx = f dx^1 \wedge dx^2 \wedge ... \wedge dx^n$ for some
function $f: U \to {\mathbf R}$ where $U$ is an open subset of $M$ on which the
$x^i$'s are defined).

Define the operator $D_i$ (total derivative) by $\displaystyle D_i =
\frac{\partial}{\partial x^i} + u^a_{iJ}\frac{\partial} {\partial u^a_J}$ (note
that we use the summation convention, i.e., the sum is over all $a$ and
multi-index $J$), and recall that the Euler-Lagrange operator maps $\Omega
^{n,0}(J^\infty E)$ into $\Omega ^{n,1}(J^\infty E)$ and is defined by $$\E (P
\nu)=\E _a(P)(\theta ^a \wedge \nu)$$ where $P \in Loc_E, \nu$ is a volume
element on the base manifold $M$, and the components $\E_a(P)$ are given by
$$\E_a(P)=(-D)_I(\frac{\partial P} {\partial u^a_I}).$$ For simplicity of
notation we will frequently use $\E(P)$ for $\E(P\nu)$. We will also use
$\tilde{D}_i$ for $\displaystyle \frac{\partial}{\partial \tilde{x}^i} +
\tilde{u}^a_{iJ}\frac{\partial}{\partial \tilde{u}^a_J}$ and $\tilde{\E}_a(P)$
for $\displaystyle (-\tilde{D})_I(\frac {\partial P} {\partial \tilde{u}^a_I})$
so that $\E(P) = \tilde{\E}_a(P) (\tilde{\theta} ^a \wedge \nu)$ in the
$(\tilde{x}^{\mu},\tilde{u}^a)$ coordinate system.

Let $\Gamma _c E$ be the subspace of $\Gamma E$ such that $\phi \in \Gamma _c
E$ iff $\phi$ is of compact support, and let $\Omega ^{k,0}_c(J^ \infty E)$ be
the subspace of $\Omega ^{k,0}(J ^\infty E)$, for $k \neq n$, such that $\alpha
\in \Omega ^{k,0}_c(J^ \infty E)$ iff $(j^\infty \phi)^* \alpha$ has compact
support for all $\phi \in \Gamma _c E$. Let $\Omega ^{n,0}_{c}(J^ \infty E)$ be
the subspace of $\Omega ^{n,0}(J^ \infty E)$ such that $P\nu \in \Omega
^{n,0}_{c} (J^ \infty E)$ iff $(j^\infty \phi)^*(P\nu)$ and $(j^\infty \phi)^*
\E_a(P)$ have compact support for all $\phi \in \Gamma _c E$ and for all $a$.
We are interested in the ``de Rham complex" $$ 0 \to \Omega ^{0,0}_c(J^\infty
E) \to \Omega ^{1,0}_c(J^\infty E) \to \cdots \to \Omega ^{n-1,0}_c(J^\infty E)
\to \Omega ^{n,0}_{c}(J^\infty E)$$ with the differential $d_H$ defined by $d_H
= dx^i D_i$, i.e., if $\alpha = \alpha _I dx^I$ then $d_H \alpha = D_i \alpha
_I dx^i \wedge dx^I$. We assume that this complex is {\it exact} which may
require the vector bundle $\pi: E \to M$ to be trivial (also see \cite{A96},
\cite{BT82}, and \cite{S89}).

Now let $\mathcal F$ be the space of local functionals where $\mathcal P \in
\mathcal F$ iff $\displaystyle \mathcal P=\int_M P \nu$ for some $P \in
Loc_E^0$, and define a Poisson bracket on $\mathcal F$ by $$\{\mathcal
P,\mathcal Q\}(\phi)= \int_M [\omega (\E(P),\E(Q))\circ j^\infty \phi] \nu,$$
where $\phi \in \Gamma _c E$, $\nu$ is a volume on $\displaystyle M, \mathcal
P= \int_M P \nu, \mathcal Q=\int_M Q \nu,$ and $P, Q \in Loc_E^0$. Using local
coordinates $(x^\mu,u^a_I)$ on $J^\infty E$, observe that for $\phi \in \Gamma
E$ such that the support of $\phi$ lies in the domain $\Omega$ of some chart
$x$ of $M$, one has $$\{\mathcal P,\mathcal Q\}(\phi) = \int_{x(\Omega)}
([\omega ^{ab} \E_a(P)\E_b(Q)] \circ j^\infty\phi \circ x^{-1})
(x^{-1})^*(\nu)$$ where $x^{-1}$ is the inverse of $x=(x^\mu)$.

{\em We assume that $\omega$ satisfies the necessary conditions for the above
bracket to satisfy the Jacobi identity, e.g. see \cite{O86} section 7.1.}

The functions $P$ and $Q$ in our definition of the Poisson bracket of local
functionals are representatives of $\mathcal P$ and $\mathcal Q$ respectively,
since generally these are not unique. In fact $\mathcal F \simeq H^n_c(J^\infty
E)$, where $H^n_c(J^\infty E) = \Omega ^{n,0}_{c}(J^\infty E)/(\textup{im}
d_H\bigcap\Omega ^{n,0}_{c}(J^\infty E))$ and im$d_H$ is the image of the
differential $d_H$.

Let $\psi:E\rightarrow E$ be an automorphism, sending fibers to fibers, and let
$\psi _M:M\rightarrow M$ be the induced diffeomorphism of $M$. Notice that
$\psi$ induces an automorphism $j^\infty\psi: J^{\infty}E \rightarrow
J^{\infty}E$ where $$(j^\infty \psi)((j^\infty \phi)(p))=j^\infty (\psi \circ
\phi \circ \psi ^{-1}_M)(\psi _M(p)),$$ for all $\phi \in \Gamma E$ and all $p$
in the domain of $\phi$.  Local coordinate representatives of $\psi _M$ and
$j^\infty \psi$ may be described in terms of charts $(\Omega,x)$ and
$(\tilde{\Omega},\tilde{x})$ of $M$, and induced charts $((\pi ^\infty)^{-1}
(\Omega),(x^{\mu},u^a))$ and $((\pi ^\infty)^{-1}
(\tilde{\Omega}),(\tilde{x}^{\mu},\tilde{u}^a))$ of $J^ \infty E$. In these
coordinates the independent variables transform via $\tilde{x}^\mu = \psi
_M^\mu(x^\nu)$.

\begin{remark} In section 4 we will consider (left) Lie group
actions on $E$ and their induced (left) actions on $J^ \infty E.$ Such actions
are defined by homomorphisms from the group into the group of automorphisms of
$E.$ \end{remark}

\begin{definition} $\omega:\Omega ^{n,1}_0 \times \Omega ^{n,1}_0
\rightarrow
Loc_E$ is {\em covariant} with respect to an automorphism $\psi:E
\rightarrow E$
of the above form iff $$\omega((j^\infty \psi)^*
\theta,(j^\infty\psi)^* \theta ') = (\textup{det}\psi _M)(j^\infty\psi)^*
(\omega(\theta,\theta ')),$$
for all $\theta,\theta ' \in \Omega ^{n,1}_0(J^ \infty E)$.
\end{definition}

\subsection{Automorphisms preserving the Poisson structure}
Let $L:J^{\infty}E \rightarrow {\bf R} \in Loc_E$ be a Lagrangian where
generally we will assume that any element of $Loc_E$ is a Lagrangian.
Let $\hat{\psi}$ denote the mapping representing the
action of the automorphism
on sections of $\pi$, i.e. $\hat{\psi}: \Gamma E \rightarrow \Gamma E$ where
$\hat{\psi}(\phi) = \psi \circ \phi \circ \psi ^{-1}_M$ and $\phi$ is a
section
of $\pi$.
This induces a mapping on the space of local functionals given by
\begin{eqnarray*}
(\mathcal P \circ \hat{\psi})(\phi)
& = &
\mathcal P (\psi \circ \phi \circ \psi ^{-1}_M) \\ & = &
\int_M [P\circ j^\infty (\psi \circ \phi \circ \psi ^{-1}_M)] \nu \\ & = &
\int_M [P\circ j^\infty\psi \circ j^\infty \phi\circ\psi ^{-1}_M)]\nu \\ & = &
\int_M [P\circ j^\infty\psi \circ j^\infty \phi] (\textup{det} \psi _M) \nu, \\
\end{eqnarray*}
where $$\mathcal P (\phi)=\int_M (P \circ j^\infty \phi) \nu,$$ and $\phi$ is a
section
of $\pi$.
We quote the following from \cite{AF03}.
\begin{theorem} \label {canonical} 
Let $\psi:E \rightarrow E$ be an automorphism of $E$ sending fibers to
fibers,
and let $\Psi: {\mathcal F} \rightarrow {\mathcal F}$ be the induced
mapping defined by $\Psi({\mathcal P})
= \mathcal P \circ \hat{\psi}$ where $\hat{\psi}:\Gamma E \rightarrow \Gamma E$
is given
by $\hat{\psi}(\phi) = \psi \circ \phi \circ \psi _M^{-1}$. Then $\Psi$ is
canonical in the sense that $$\{\Psi(\mathcal P),\Psi(\mathcal
Q)\}=\Psi(\{\mathcal P,\mathcal Q\}),$$ for all $\mathcal P, \mathcal Q
\in
\mathcal F$ iff $\omega$ is covariant with respect to $\psi$.
\end{theorem}

\section{Canonical automorphisms and sh-Lie algebras} In this
section we consider the structure maps of the sh-Lie algebra on the horizontal
complex \{$\Omega ^{i,0}(J^\infty E)$\} where we assume that the vector bundle
$E$ is {\em trivial}. A detailed study of sh-Lie algebras can be found in
\cite{BFLS98} or one of the references therein; however it is useful to give a
brief overview.

\subsection{Overview of sh-Lie algebras} \label{osh} Let ${\cal F}$ be a vector
space and $(X_*,l_1)$ a homological resolution thereof, i.e., $X_*$ is a graded
vector space, $l_1$ is a differential and lowers the grading by one with ${\cal
F}\simeq H_0(l_1)$ and $H_k(l_1)=0$ for $k>0$. The complex $(X_*,l_1)$ is
called the resolution space. (We are {\it not} using the term `resolution' in a
categorical sense.) Consider a homological resolution of the space ${\cal F}$
of local functionals as in \cite{BFLS98}. In the field theoretic framework
considered in \cite{BFLS98} it was shown that under certain hypothesis (see the
Theorem below) the Lie structure defined by the Poisson bracket on ${\cal F}$
induces an sh-Lie structure on the graded vector space
$X_i=\Omega^{n-i,0}(J^{\infty}E),$ for $0\leq i \leq n$. For completeness we
give the definition of sh-Lie algebras and include a statement of the relevant
theorem.

\begin{definition}
An sh-Lie structure on a graded vector space $X_*$ is a collection of
linear,
skew-symmetric maps $l_k: \bigotimes ^k X_* \to X_*$ of degree $k-2$ that
satisfy
the relation
$$\sum_{i+j=n+1}\sum_{unsh(i,n-i)} e(\sigma)(-1)^ \sigma
(-1)^{i(j-1)}l_j(l_i
(x_{\sigma(1)},\cdots,x_{\sigma(i)}),\cdots,x_{\sigma(n)}) = 0,$$ where $
1
\leq i,j$.
\end{definition}
Notice that in this definition $e(\sigma)$ is the Koszul sign which depends
on the permutation $\sigma$ as well as on the degree of the elements
$x_1,x_2,\cdots,x_n$ (a minus sign is introduced whenever two
consecutive odd elements are permuted, see for example \cite{LM95}).

\begin{remark} Although this may seem to be a rather complicated structure, it
simplifies drastically in the case of field theory where, aside from the
differential $l_1=d_H,$ the only non-zero maps are $l_2$ and $l_3$ in degree 0.
\end{remark}

The Theorem relevant to field theory depends on the existence of a linear
skew-symmetric map $\tilde{l}_2: X_0 \otimes X_0 \to X_0$ (in our case the
Poisson bracket will be the integral of this mapping as we will see in detail
shortly) satisfying conditions $(i)$ and $(ii)$ below. These conditions are all
that is needed in order that an sh-Lie structure exists.

\begin{theorem} \label{shLie}
A skew-symmetric linear map $\tilde{l}_2: X_0 \otimes X_0 \to X_0$ that
satisfies
conditions (i) and (ii) below extends to an sh-Lie structure on the
resolution space $X_*$;\\
$(i) \quad \tilde{l}_2(c,b_1) = b_2$ \\
$\displaystyle (ii) \sum _{\sigma \in unsh(2,1)} (-1)^\sigma
\tilde{l}_2(\tilde{l}_2(c_{\sigma(1)},c_{\sigma(2)}),c_{\sigma(3)}) =b_3$
\\
where $c, c_1, c_2, c_3$ are cycles and $b_1, b_2, b_3$ are boundaries in
$X_0$.
\end{theorem}

\subsection{The effect of a canonical automorphism on the structure maps
of the sh-Lie algebra}
We find it convenient here to quote the main result from \cite{AF03}.
First define $\tilde l_2$ on
$\Omega ^{n,0} \otimes \Omega ^{n,0}$ by
\begin{equation} \label{l2}
\tilde{l}_2(P\nu ,Q\nu)= \omega ^{ab}\E_b(Q)\E_a(P) \nu
=\omega(\E(P),\E(Q)) \nu.
\end{equation}

\noindent We have
\begin{theorem} \label{invl2} Let $\psi:E \rightarrow E$ be
an automorphism of $E$ sending fibers to fibers and recall that $j^\infty \psi$
is the induced automorphism on $J^\infty E$. We have $$\tilde{l}_2((j^\infty
\psi)^* \alpha,(j^\infty \psi)^* \beta) = (j^\infty
\psi)^*(\tilde{l}_2(\alpha,\beta))$$ for all $\alpha,\beta \in
\Omega^{n,0}(J^{\infty}E)$ iff $\omega$ is covariant with respect to $\psi$.
Moreover
$$l_3((j^\infty\psi)^*\alpha,(j^\infty\psi)^*\beta,(j^\infty\psi)^*\gamma) =
(j^\infty\psi)^*l_3(\alpha,\beta,\gamma) + l_1(\delta),$$ for all
$\alpha,\beta,\gamma \in \Omega^{n,0}(J^{\infty}E)$, and for some $\delta \in
\Omega^{n-2,0}(J^{\infty}E)$. \end{theorem}

Now we are ready to consider the symmetry reduction ideas and apply them to
define an (induced) sh-Lie structure on a reduced complex.

\section{Symmetry reduction of the graded vector space} 

Let $\pi: E \rightarrow M$ be a vector bundle and let $J^\infty E$ be the
infinite jet bundle of $E$ as before. Suppose that $G$ is a Lie group acting on
$E$ via automorphisms (as in section $\ref{BAS}$) and hence inducing an action
of $G$ on $J^\infty E$. We assume the induced action $\hat{\psi}_g$ on $\Gamma
E$ is canonical with respect to the Poisson bracket of local functionals for
all $g \in G$. Notice that $G$ acts via canonical tranformations on the space
of local functionals if and only if for every $j^\infty \psi _g$
$$\tilde{l}_2((j^\infty \psi _g)^* f_1,(j^\infty \psi _g)^* f_2)= (j^\infty
\psi _g)^*(\tilde{l}_2(f_1,f_2)),$$ where $\tilde{l}_2$ is defined on the
vector space $\Omega^{n,0}(J^\infty E)$ as in the previous section (in fact
$\displaystyle \tilde{l}_2(f_1,f_2) = \frac{1}{2} [\omega(\E(f_1),\E(f_2)) -
\omega(\E(f_2),\E(f_1))]$, see also equation $\ref {l2}$).

We use the ideas of symmetry reduction as in \cite{AF97} which
lead to a cochain map between cochain complexes. Specifically we find a cochain
map between the variational bicomplexes of $J^\infty E$ and $J^\infty
\overline{E}$ where $\overline{E} = E/G$ and $G$ is the Lie group acting on $E$.
After that we show how one gets an sh-Lie structure on a reduced graded vector
space (complex).

\subsection{Symmetry reduction}

We begin with an introduction of the basic ideas of symmetry reduction and refer
the reader to \cite{AF97} for more details.
Suppose that $G$ is a $p$-dimensional Lie group acting on $E$ and inducing an
action on $M$. Also suppose that the dimension of $M$ is $n$ and that of $E$ is
$n+m$. We assume that $G$ acts transversally to the fibers of $E$, and
that it acts projectably on $\pi:E\to M$, and regularly and effectively on both
$E$ and $M$ with orbits of
dimension $q < n$. It follows that the quotient spaces $\overline{E}=E/G$ and
$\overline{M}=M/G$ are smooth manifolds of dimensions $n+m-q$ and $n-q$
respectively, and that the following diagram commutes with all smooth maps.

\[ \begin{array}{rcccl}
& E & \xrightarrow{\pi _{\overline{E}}} & \overline{E} & \\
\pi& \downarrow & & \downarrow &\overline{\pi} \\
& M & \xrightarrow{\pi _{\overline{M}}} & \overline{M} &
\end{array} \]

On $E$ one uses local coordinates $(x^i,u^a)$ and local coordinates
$(\hat{x}^i, y^r, v^\alpha)$ adapted to $G$ such that the locally $G$-invariant
sections (as defined below) of
$\pi:E\to M$ are given by $v^\alpha = f(y^r)$. So, if for example, $G=SO(3)$
acts on $M={\mathbf R}^3-\{0\}$ and $E=M\times {\mathbf R}^2$ then
$(x,y,z,u^1,u^2)$
are coordinates on $E$ while $$\hat{x}=x \quad \hat{y}=y \quad
r=\sqrt{x^2+y^2+z^2} \quad v^1=u^1 \quad v^2=u^2$$ are coordinates on an open
subset of $E$
adapted to $G$ where the last three, namely $(r, v^1, v^2)$, can
serve as coordinates on $\overline{E}$.

Let $J^\infty \overline{E}$ denote the infinite jet bundle of
$\overline{E}$, and let $\Omega _{\textup{pr}G}^{r,s} (J^\infty E)$ denote the
subspace of $\Omega ^{r,s} (J^\infty E)$ that consists of the forms that are
invariant under the prolonged action of $G$.

Let $\mathfrak{g}$ denote the Lie algebra of $G$, and let $\Gamma$ denote the
vector space of infinitesimal generators of the action of $G$ on $E$. A
differential form $\gamma$ on $J^\infty E$ is an invariant of the action if
${\mathcal L}_{\textup{pr} X} \gamma = 0$ for all $X \in \Gamma$, and where
pr$X$ is the prolongation of $X$ to $J^\infty E$.

A vector $X_\sigma$ at $\sigma = j^\infty(s)(p) \in J^\infty E$, where $s$ is a
local section of the bundle $\pi: E\to M$, is called a total vector at $\sigma$
if
$$X_\sigma (f) = [(\pi ^\infty _M)_*(X_\sigma)](f \circ j^\infty(s))$$ for all
$f:U \to {\mathbf R}$ where $U$ is any subset of $J^\infty E$ containing
$\sigma$, and where $\pi ^\infty _M:J^\infty E \to M$ is the canonical
projection. We denote the space of all total vector fields on $J^\infty E$ by
Tot($J^\infty E$). If $X$ is a vector field on $M$ then tot$X$ denotes
the extension of $X$ into a total vector field on $J^\infty E$. In local
coordinates total vectors are spanned by the total derivatives
$\displaystyle D_i = \frac{\partial}{\partial x^i} + u^a_{iJ}\frac{\partial}
{\partial u^a_J}, i= 1, 2, \cdots, n.$

A section $s$ of $\pi:E\to M$ (here we use $s$ for sections rather than $\phi$
which was used earlier in the paper)
is locally $G$-invariant if for all $g \in G$
sufficiently close to the identity $g\cdot [s(g^{-1}\cdot p)] = s(p)$. The jet
space of $G$-invariant local sections of $E$ is the bundle
$\textup{Inv}_G^\infty (E) \to M$ defined by
\[ \begin{array}{cl}
\textup{Inv}_G^\infty (E) = \{
\sigma \in J^\infty E | \sigma = j^\infty(s)(p), & s
\textup{ is a locally } \\
& G\textup{-invariant section of } E\}.
\end{array} \]

In local coordinates $(\hat{x}^i, y^r,v^\alpha)$ of $E$ adapted to $G$
the locally $G$-invariant sections of $\pi:E\to M$ are
given by $v^\alpha = f(y^r)$ and therefore $$\textup{Inv}_G^\infty (E) = \{
\sigma =(\hat{x}^i, y^r, v^\alpha, v^\alpha _i, v^\alpha _r, v^\alpha _{ij},
v^\alpha _{ir},v^\alpha _{rs}, \cdots) | v^\alpha _i=0, v^\alpha _{ij}=0,
v^\alpha _{ir}=0, \cdots\}.$$

If $s:M\to E$ is a $G$-invariant local section then there exists a unique local
section of $\overline{\pi}$ such that
\begin{equation} \label {secred}
\overline{s}(\pi _{\overline{M}} (p)) = \pi _{\overline{E}}(s(p)).
\end{equation}
This correspondence between $G$-invariant local sections of $\pi$ and local
sections of $\overline{\pi}$
induces a projection map $\Pi: \textup{Inv}_G^\infty (E) \to J^\infty
\overline{E}$ defined by $\Pi (j^\infty(s)(p)) = j^\infty(\overline{s})(\pi
_{\overline{M}} (p))$, see \cite{AF97}. \\

One can describe the correspondence between $G$-invariant objects on $\pi:E\to
M$ and the associated objects on $\overline{\pi}:\overline{E} \to \overline{M}$.
If $s:M\to E$ is $G$-invariant then we define $\overline{s}=\varrho(s)$ to be
the unique section of $\overline{\pi}$ satisfying $\ref{secred}$.

A $G$-invariant form $\alpha$ on $E$ satisfying $X \lh \alpha = 0$ for all
$X$ in $\Gamma$ is said to be $G$-{\em basic}. If $\alpha$ is $G$-basic then
there exists a unique form $\overline{\alpha}=\varrho(\alpha)$ on $\overline{E}$
such that $$\alpha = (\pi _{\overline{E}})^*(\overline{\alpha}).$$
If $f:J^\infty E \to {\mathbf R}$ is a $G$-invariant function then there is a
unique function $\overline{f}:J^\infty \overline{E} \to {\mathbf R}$ satisfying
$\overline{f}(\overline{\sigma})=f(\sigma)$ where $\overline{\sigma} \in
J^\infty \overline{E}, \sigma \in \textup{Inv}_G^\infty (E)$ and $\Pi(\sigma) =
\overline{\sigma}$. We let $\varrho(f)=\overline{f}$.
More generally if $\alpha \in \Omega _{\textup{pr}G}^{r,s} (J^\infty E)$ and
$\textup{tot}X \lh \alpha = 0$ for all $X$ in $\Gamma$, then there is a unique
$\overline{\alpha} \in \Omega ^{r,s} (J^\infty \overline{E})$ such that $$\imath
^*(\alpha)=\Pi ^*(\overline{\alpha})$$ where $\imath:\textup{Inv}_G^\infty (E)
\to J^\infty E$ is the canonical inclusion (see \cite{AF97}). We let
$\varrho(\alpha)=\overline{\alpha}$.

\begin{example}
Suppose that $f:J^\infty E \to {\mathbf R}$ is a $G$-invariant function
expressed in local coordinates adapted to $G$ by $f(\hat{x}^i, y^r, v^\alpha,
v^\alpha _i, v^\alpha _r, v^\alpha _{ij},
v^\alpha _{ir},v^\alpha _{rs}, \cdots)$ then $$\varrho (f)(y^r,
v^\alpha, v^\alpha _r, v^\alpha _{rs}, \cdots) = f(\hat{x}^i, y^r,
v^\alpha,0, v^\alpha _r, 0,0,v^\alpha _{rs}, \cdots).$$
As an illustration suppose that $G=SO(3)$
acts on $M={\mathbf R}^3-\{0\}$ and that $E=M\times {\mathbf R}$. Let $f=u_{xx}+
u_{yy} + u_{zz}$. Using local coordinates $(x,y,z,u)$
on $E$ and local coordinates adapted to $G$ $$\hat{x}=x \quad \hat{y}=y \quad
r=\sqrt{x^2+y^2+z^2} \quad v=u,$$ one calculates (by the chain rule)
$\displaystyle u_x=u_{\hat{x}} + \frac{x}{r} u_r, u_y=u_{\hat{y}} + \frac{y}{r}
u_r,$
and $$u_{xx}=v_{\hat{x}\hat{x}} + \frac{x}{r}v_{\hat{x}r} +
\frac{x}{r}(v_{r\hat{x}} + \frac{x}{r}v_{rr}) + v_r \frac{r^2-x^2}{r^3},$$
$$u_{yy}=v_{\hat{y}\hat{y}} + \frac{y}{r}v_{\hat{y}r} +
\frac{y}{r}(v_{r\hat{y}} + \frac{y}{r}v_{rr}) + v_r \frac{r^2-y^2}{r^3},$$
$$u_{zz} =
\frac{z^2}{r^2}v_{rr} + \frac{r^2-z^2}{r^3} v_r.$$
Now set $v_{\hat{x}\hat{x}}=v_{\hat{x}r}=v_{r\hat{x}}=0$ and
$v_{\hat{y}\hat{y}}=v_{\hat{y}r}=v_{r\hat{y}}=0$
to get
$\boxed{\frac{x^2}{r^2}v_{rr} + \frac{r^2-x^2}{r^3} v_r}$ for $u_{xx}$ and
$\boxed{\frac{y^2}{r^2}v_{rr} + \frac{r^2-y^2}{r^3} v_r}$ for $u_{yy}$ so that
$\displaystyle \varrho (f) = v_{rr} + \frac{2}{r} v_r$. \qed
\end{example}

\begin{example}
Let $\alpha
\in \Omega _{\textup{pr}G}^{r,0} (J^\infty E)$ be $G$-invariant and suppose that
$\textup{tot}X \lh \alpha = 0$ for all $X$ in $\Gamma$ then $\alpha$ can be
expressed in local coordinates adapted to $G$ as $\alpha= A_{i_1i_2\cdots i_r}
(dy^{i_1} \wedge dy^{i_2}\wedge \cdots \wedge dy^{i_r})$ where the
$A_{i_1i_2\cdots i_r}$'s are $G$-invariant, and $\varrho (\alpha) = \varrho
(A_{i_1i_2\cdots i_r}) (dy^{i_1} \wedge dy^{i_2}\wedge\cdots \wedge dy^{i_r})$.
\qed
\end{example}

Finally recall that a $q$ multi-vector on $M$ is an alternating tensor of
type $(q,0)$. Let $\Gamma _M$ be a Lie algebra of vector fields on
$M$, then a {\em $q$-chain} ${\mathcal X}$ on $\Gamma _M$ is a (non-zero) $q$
multi-vector that can be expressed as
${\mathcal X} = J (X_1 \wedge X_2 \wedge \cdots \wedge X_q),$
where $J$ is a function on $M$ and $X_1, X_2, \cdots, X_q \in \Gamma _M$. \\

Now we show how a cochain map between the cochain ``de Rham complexes" of $E$
and $\overline{E}$ may be defined. Recall that $\Gamma$ denotes the {\em vector
space of infinitesimal generators} of the action of $G$ on $E$. One assumes the
existence of a $G$-invariant $q$-chain on Tot $\Gamma$ $${\mathcal X} = J(
\textup{tot}X_1 \wedge \textup{tot}X_2 \wedge \cdots \wedge \textup{tot}X_q),$$
where $J$ is a function on $J^\infty E$ and $X_1, X_2, \cdots, X_q \in \Gamma$,
such that the map $\varrho _{\mathcal X}: \Omega _{\textup{pr}G}^{r,s}
(J^\infty E) \to \Omega ^{r-q,s} (J^\infty \overline{E})$ defined by $$\varrho
_{\mathcal X}(\gamma) = (-1)^{q(r+s)} \varrho ({\mathcal X}\lh \gamma)$$ is a
$d_H$ cochain map between cochain complexes, and such that  $$\E(\varrho
_{\mathcal X} \delta) = \varrho _{\mathcal X}(\E(\delta))$$ for all $\delta \in
\Omega _{\textup{pr}G}^{n,0} (J^\infty E)$. In \cite{AF97} it was proved that
if the action of $G$ on $E$ is free and $G$ is unimodular (i.e. $\displaystyle
\sum _{a=1} ^q C^a_{ia} = 0$ where the $C^a_{bc}$'s are the structure constants
of $\mathfrak{g}$) then the existence of the $G$-invariant $q$-chain follows.
In fact in \cite{AF97} the existence of such $q$-chains is studied in detail.
Notice that the map $\varrho _{\mathcal X}$ is onto but generally {\it not}
one-to-one.

\begin{remark} The above correspondence $\varrho$ can be restricted to
compact-support subcomplexes as defined earlier in this paper, in particular to
the first row of such subcomplexes (the row consisting of purely horizontal
forms). If $s:M\to E$ is a $G$-invariant section of compact support then
$\overline{s}=\varrho(s):\overline{M}\to \overline{E}$ is of compact
support. If $\alpha \in \Omega _{\textup{pr}G}^{r,0} (J^\infty E)$
satisfies $\textup{tot}X \lh \alpha = 0$ for all $X$ in $\Gamma$, and
$(j^\infty s)^*(\alpha)$ is of compact support for all sections $s:M\to E$ of
compact support then for (the unique form) $\overline{\alpha} = \varrho(
\alpha) \in \Omega ^{r,0} (J^\infty \overline{E})$ we have: $(j^\infty
\overline{s})^*(\overline{\alpha})$ is of compact support for all sections
$\overline{s}:\overline{M}\to \overline{E}$ of compact support, ...etc.
\end{remark}

Henceforth we will work with the compact-support subcomplexes that consist of
purely horizontal forms, where $\varrho _{\mathcal X}$ serves as a $d_H$
cochain map between these subcomplexes (though some of our conclusions apply to
the more general complexes).

\subsection{sh-Lie structure on the reduced graded vector space}

We assume that $(\Omega ^{*,0}_c(J^\infty E),d_H)$ and $(\Omega
^{*,0}_c(J^\infty \overline{E}),d_H)$ are exact. The map $\tilde{l}_2$ on
$\Omega ^{n,0}_c(J^\infty E)$ induces a map $\hat{l}_2$ on $\Omega
^{n-q,0}_c(J^\infty \overline{E})$ as follows. Define $\hat{l}_2:\Omega
^{n-q,0}_c(J^\infty \overline{E})\otimes \Omega ^{n-q,0}_c (J^\infty
\overline{E}) \to \Omega ^{n-q,0}_c(J^\infty \overline{E})$ by
$$\hat{l}_2 (\varrho_{\mathcal X} \alpha,\varrho_{\mathcal X} \beta) =
\varrho_{\mathcal X} (\tilde{l}_2 (\alpha, \beta))$$
for all $G$-invariant $\alpha,\beta$ in $\Omega ^{n,0}_c(J^\infty E)$.
Now the question is: Is the map $\hat{l}_2$
well-defined? \\ \indent First recall that the cochain map $\varrho_{\mathcal
X}$ is onto. Now assume that $\omega$ is covariant with respect to the group
action of $G$ (i.e.
covariant with respect to $\psi _g$ for all $g \in G$ as in section $\ref
{osh}$) so that by Theorem $\ref{invl2}$ it follows that if $\alpha$ and
$\beta$ are in  $\Omega _{\textup{pr}G}^{n,0} (J^\infty E)$ then so is
$\tilde{l}_2 (\alpha, \beta)$. Finally we assume that if $\varrho_{\mathcal X}
\alpha = \varrho_{\mathcal X} \alpha '$ for any $\alpha, \alpha ' \in \Omega
_{\textup{pr}G}^{n,0} (J^\infty E)$ then $\hat{l}_2 (\varrho_{\mathcal X}
\alpha,\varrho_{\mathcal X} \beta) - \hat{l}_2 (\varrho_{\mathcal X} \alpha
',\varrho_{\mathcal X} \beta) = \varrho_{\mathcal X}\tilde{l}_2 (\alpha -
\alpha ', \beta) = \varrho_{\mathcal X}[\omega (\E(\alpha - \alpha
'),\E(\beta)) \nu] = 0$, so that $\hat{l}_2$ is {\em well-defined}.

Observe that if the volume form $\nu$ is $G$-invariant then $\hat{l}_2$ is
given by
$$\hat{l}_2 (\varrho_{\mathcal X} \alpha,\varrho_{\mathcal X} \beta)=
\overline{\omega}(\E(\varrho_{\mathcal X} \alpha),\E(\varrho_{\mathcal X}
\beta)) \overline{\nu}$$
where $\overline{\omega}:
\Omega ^{n-q,1}_0(J^\infty \overline{E}) \times \Omega ^{n-q,1}_0(J^\infty
\overline{E}) \to Loc_{\overline{E}}$ is
defined by
$$\overline{\omega}
(\varrho_{\mathcal X}\gamma,\varrho_{\mathcal X}\delta) := \varrho
(\omega(\gamma,\delta))$$ for $\gamma,\delta$ that lie in the image of $\E$ in
$\Omega _{\textup{pr}G}^{n,1}(J^\infty E)$. Thus
$\overline{\omega}(\E(\varrho_{\mathcal X} \alpha),\E(\varrho_{\mathcal X}
\beta))=\overline{\omega}
(\varrho_{\mathcal X}\E(\alpha),\varrho_{\mathcal X}\E(\beta)) = \varrho
[\omega(\E(\alpha),\E(\beta))]$. Here $\overline{\nu}$ satisfies
$(\overline{\pi}_M^\infty)^*\overline{\nu}=\varrho_{\mathcal X}
(\pi _M^\infty)^*\nu$ where
$\overline{\pi}_M^\infty:J^\infty \overline{E}\to \overline{M}$ and
$\pi _M^\infty:J^\infty E \to M$ are the usual projections of the corresponding
jet bundles.

Skew-symmetry and linearity of
$\hat{l}_2$ follow easily from the skew-symmetry and linearity of
$\tilde{l}_2$ in addition to the linearity of $\varrho_{\mathcal X}$.
Furthermore $\hat{l}_2$ satisfies \\

$(i) \quad \quad \hat{l}_2(d_H k_1,h) = d_H k_2,$ \\ \indent
$\displaystyle (ii) \quad \sum _{\sigma \in unsh(2,1)} (-1)^\sigma
\hat{l}_2(\hat{l}_2(f_{\sigma(1)},f_{\sigma(2)}),f_{\sigma(3)}) =d_H
k_3,$

\noindent
for all $k_1 \in
\Omega ^{n-q-1,0}_c(J^ \infty \overline{E}), h,f_1,f_2,f_3 \in \Omega
^{n-q,0}_c(J^\infty \overline{E})$ and
for some $k_2,
k_3 \in \Omega ^{n-q-1,0}_c(J^ \infty \overline{E})$. {\em Subsequently we will
suppress some of the notation and assume
the summands are over the appropriate shuffles with their corresponding
signs}.
\\ \indent
Notice that $(i)$ follows in the strong sense $\hat{l}_2(d_H k_1,h) = 0$ since
$\E(d_H k_1)=0$. While to verify $(ii)$, let $f_i=\varrho_{\mathcal X} F_i$ for
$i=1,2,3$, and where $f_1,f_2,f_3 \in \Omega^{n-q,0}_c(J^\infty \overline{E})$
are arbitrary, and notice that 
\begin{eqnarray*}
\displaystyle \sum _{\sigma}
\hat{l}_2(\hat{l}_2(f_{\sigma(1)},f_{\sigma(2)}),f_{\sigma(3)})
& = &
\varrho_{\mathcal X} \sum _{\sigma}\tilde{l}_2(\tilde{l}_2(
F_{\sigma(1)},F_{\sigma(2)}),F_{\sigma(3)}) \\
& = &
\varrho_{\mathcal X} (d_H K_2) \\
& = &
d_H (\varrho_{\mathcal X} K_2),
\end{eqnarray*}
since $\varrho_{\mathcal X}$ is a $d_H$ cochain map and
where the sum is over the unshuffles (2,1), and for some
$K_2 \in \Omega^{n-1,0}_{\textup{pr}G}(J^\infty E)$. We have shown:

\begin{theorem}
The skew-symmetric linear map $\hat{l}_2$ as defined above on the space
$\Omega ^{n-q,0}_c(J^\infty \overline{E})$ extends to an sh-Lie structure on
the exact graded space $(\Omega ^{*,0}_c(J^\infty \overline{E}), d_H)$.
\end{theorem}
\noindent
We have also shown (see lemmas 1 and 2 in \cite{BFLS98})
\begin{theorem}
There exists a skew-symmetric bilinear bracket on $\overline{H}_0 \times
\overline{H}_0$, where we use $\overline{H}_0$ for
$H^{n-q}(\Omega ^{*,0}_c (J^\infty \overline{E}),d_H)$,
that satisfies the Jacobi identity. This bracket is
induced by the map $\hat{l}_2$.
\end{theorem}

In fact, if $\nu$ is $G$-invariant with $\varrho_{\mathcal X} (\pi _M^*\nu) =
\overline{\pi}_M^*\overline{\nu}$ then
this bracket can be identified with $\displaystyle \{\mathcal P,\mathcal
Q\} = \int_{\overline{M}} \overline{\omega}(\E(P),\E(Q)) \overline{\nu}$, where
$\displaystyle {\mathcal P} = \int_{\overline{M}} P\overline{\nu}
\textup{ and } {\mathcal Q} = \int_{\overline{M}}
Q \overline{\nu}$ represent arbitrary elements of $\overline{H}_0$, and
$P, Q \in Loc_{\overline{E}}^0$.

\begin{example}
Consider $M={\mathbf R}^3 -\{0\}, E=M \times {\mathbf R}^2$, and let
\[ \omega = \left( \begin{array}{cc}
0 & r \\
-r & 0 \end{array} \right), \]
where $r=\sqrt{x^2+y^2+z^2}$ and $x,y,z$ are cartesian coordinates on $M$.
Recall that $\omega$ defines the map $\tilde{l}_2$. Now
let $G=SO(3)$ act on $E$ by rotations of the base manifold $M$. The Lie algebra
of the action has
generators $$X_1= x \partial y - y \partial x, \quad 
X_2= y \partial z - z \partial y, \quad X_3= z \partial x -
x \partial z.$$
Observe that $\omega$ is covariant with respect to $G$.
In fact the components of $\omega$ are $G$-invariant in this case. Now
notice that $\displaystyle {\mathcal X}= \frac{r}{y} (\textup{tot}X_1 \wedge
\textup{tot}X_2)$ is a $G$-invariant
2-chain, and that $\varrho_{\mathcal X} (dx\wedge dy\wedge dz) = r^2 dr$, where
$\nu=dx\wedge dy\wedge dz$ is $G$-invariant and $\overline{\nu}= r^2 dr$.
We conclude that there exists of a reduced sh-Lie structure.
Now suppose that $$P=u^1(u^2_{xx}+u^2_{yy}+u^2_{zz}) \quad \textup{and}\quad
Q=u^1u^2,$$ then
$$\tilde{l}_2(P\nu, Q\nu)=
r [(u^2_{xx}+u^2_{yy}+u^2_{zz})u^1 - (u^1_{xx}+u^1_{yy}+u^1_{zz})u^2]\nu,$$
while $$\varrho_{\mathcal X}(P\nu)=u^1(u^2_{rr} + \frac{1}{r} u^2_r) 
\overline{\nu} \quad , \quad
\varrho_{\mathcal X}(Q\nu) = u^1u^2 \overline{\nu},$$
and
\begin{eqnarray*}
\hat{l}_2(\varrho_{\mathcal X}(P\nu),\varrho_{\mathcal X}(Q\nu))
& = &
r[(u^2_{rr} + \frac{2}{r} u^2_r) u^1 -
(u^1_{rr} + \frac{2}{r} u^1_r) u^2] \overline{\nu} \\
& = &
[(ru^2_{rr} + 2u^2_r) u^1 -
(ru^1_{rr} + 2u^1_r) u^2] \overline{\nu}
\end{eqnarray*}
\noindent
where we have used $\hat{l}_2 (\varrho_{\mathcal X}(P\nu),
\varrho_{\mathcal X}(Q\nu)) =
\varrho_{\mathcal X} (\tilde{l}_2 (P\nu,Q\nu))$ which was defined earlier.

Now notice that if for example one has $$\int_M \tilde{l}_2(P\nu, Q\nu) =
\int_{\bf R^3} r [(u^2_{xx}+u^2_{yy}+u^2_{zz})u^1 -
(u^1_{xx}+u^1_{yy}+u^1_{zz})u^2] dx \wedge dy \wedge dz$$
and one is interested
in sections that are $G$-invariant, i.e., sections that depend only on $r$, then
the above
integral reduces to $$ 4\pi \int_0^\infty r^3 [(u^2_{rr} + \frac{2}{r} u^2_r)u^1
- (u^1_{rr} + \frac{2}{r} u^1_r) u^2]
dr = 4\pi \int_{\overline{M}} \varrho_{\mathcal X}\tilde{l}_2(P\nu, Q\nu),$$
where notice that the $4 \pi$ is obtained by integrating out the variables which
the fields/sections do not depend on.

More generally, if $P \nu$ is $G$-invariant
and one has $\displaystyle \int_{\bf R^3} P \nu$ then for sections that are
$G$-invariant this integral reduces to $\displaystyle 4\pi \int_0^\infty
\varrho (P) \overline{\nu}$.
\qed
\end{example}

\section{Reduction of local functionals}
As we saw in the previous section the map $\varrho _{\mathcal X}$ defines a
correspondence between functionals on $J^\infty E$ and functionals on $J^\infty
\overline{E}$. This correspondence is ``natural" when one is interested in
$G$-invariant sections.
Given the functional $$ \int _M P \nu$$ on $J^\infty E$
where $P\nu$ is a $G$-invariant horizontal $n$-form,
its {\em reduced functional} is the functional on $J^\infty
\overline{E}$ given
by $$ \int _{\overline{M}} \varrho _{\mathcal X}(P \nu).$$
Notice that if the action of $G$ does not have a vertical component along the
fibers of $\pi:E\to M$, while $\phi$ is a $G$-invariant local section of
$\pi$ with support in $\Omega$, $\overline{\phi}$ is the
section of $\overline{\pi}: \overline{E}\to \overline{M}$ corresponding to
$\phi$, then
$$\int _\Omega (j^\infty \phi)^* P \nu = V \int _{\overline{\Omega}} (j^\infty
\overline{\phi})^*\varrho _{\mathcal X}(P \nu)$$
where $\overline{\Omega} = \pi _{\overline{M}}(\Omega)$, and
$V$ is obtained when integrating out the variables which $\phi$ does not
depend on (from the left-hand side integral).

\begin{example}
Consider the Euler equations
$$ \frac{\partial {\bf u}}{\partial t}+{\bf u} \cdot \nabla {\bf u}=-\nabla p
\quad \textup{and} \quad \nabla \cdot {\bf u} = 0,$$
where ${\bf u}= (u,v,w)$ and $p$ are the dependent variables while ${\bf x} =
(x,y,z)$ and $t$ are the independent variables. It is known that the $SO(3)$
invariant solutions are given by ${\bf u} = (a(t)/r^3) {\bf x}$ where $a(t)$ is
a function of $t$, while the energy is given by the functional
$\displaystyle {\mathcal E}(t)=\frac{1}{2} \int _{\Omega} |{\bf u} |^2 d \bf
x$ where $\Omega$ is the region of ${\bf R}^3$ over which the solution is
defined.
For an $SO(3)$ invariant solution, $\Omega$ is a spherical region and,
the energy reduces to $\displaystyle 2\pi \int _{\overline{\Omega}}
(a(t)^2/r^2)
dr$ where $\overline{\Omega}$ is the subset of ${\bf R}$ corresponding to
$\Omega$ via the action of $SO(3)$.
\end{example}

\begin{remark} In \cite{AF03} reduction of local functionals and sh-Lie
structures was studied for the case when the Lie group $G$ acts only on the
fibers, i.e., when the induced action on the base manifold $M$ is just the {\em
identity map}. In the current paper we assumed that the action is {\em
transversal} to the fibers. One may consider a general case where one has a
combination of these two kinds of action, involving two different Lie groups.
For example one can consider a reduction from $\pi$ to $\overline{\pi}$ under a
Lie group action as was done in this paper. Then another reduction on $\Omega
^{*,0}_c (J^\infty \overline{E})$ may be obtained when another Lie group acts
on the fibers of $\overline{\pi}$ as in \cite{AF03}. \end{remark}

\noindent {\bf Acknowledgements} I would like to thank Professor Ron Fulp for
very useful remarks, comments and discussions, in addition to going over drafts
of this paper.

\providecommand{\bysame}{\leavevmode\hbox to3em{\hrulefill}\thinspace}

\end{document}